\documentclass[a4paper,12pt]{article}

\usepackage{amsmath}
\usepackage{amsfonts}
\usepackage[dvips,bookmarksnumbered=true,breaklinks=true]{hyperref}
\usepackage{a4}

\newcommand{\pa}{\partial}

\newcommand{\inv}[1]{\frac{1}{#1}}
\newcommand{\starco}[2]{\left[#1\stackrel{\star}{,}#2\right]}

\newcommand{\var}[2]{\frac{\d #1}{\d #2}}

\newcommand{\intx}{\int d^4x}
\newcommand{\intk}{\int d^4k}
\newcommand{\Gam}{\Gamma^{(0)}}
\newcommand{\Act}{S}
\newcommand{\ri}{{\rm i}}
\newcommand{\ig}{{\rm i}g}

\renewcommand{\d}{\delta}
\renewcommand{\l}{\lambda}
\renewcommand{\th}{\theta}
\renewcommand{\k}{\tilde{k}}

\newcommand{\F}{\widetilde{F}}
\newcommand{\D}{\widetilde{D}}

\newcommand{\wpart}{\widetilde{\partial}}
\newcommand{\wsq}{\widetilde{\square}}
\newcommand{\bc}{\bar{c}}

\newcommand{\uim}{UV/IR mixing}
\newcommand{\nc}{non-commutative}

\title{\begin{flushright}{\small LYCEN 2008-05\\[-2.5ex]UWThPh-2008-06}\end{flushright}\vspace{1.5em}
Translation-invariant models for {\nc} gauge fields}
\author{Daniel N. Blaschke\footnotemark[1]~,
Fran\c cois Gieres\footnotemark[2]~, Erwin Kronberger\footnotemark[1]~,
\\Manfred Schweda\footnotemark[1]~ and Michael Wohlgenannt\footnotemark[3]}
\date{April 11, 2008}

\begin{document}

\maketitle

\begin{center}
\renewcommand{\thefootnote}{\fnsymbol{footnote}}
\vspace{-0.3cm}\footnotemark[1]Institute for Theoretical Physics,
Vienna University of Technology\\
Wiedner Hauptstrasse 8-10, A-1040 Vienna (Austria)\\[0.3cm]
\footnotemark[2] Universit\'e de Lyon,
Institut de Physique Nucl\'eaire,\\
Universit\'e Lyon 1 and CNRS/IN2P3, Bat. P. Dirac,\\
4 rue Enrico Fermi, F - 69622 - Villeurbanne (France)\\[0.3cm]
\footnotemark[3] Faculty of Physics, University of Vienna\\
Boltzmanngasse 5, A-1090 Vienna (Austria)\\[0.5cm]
\ttfamily{E-mail: blaschke@hep.itp.tuwien.ac.at, gieres@ipnl.in2p3.fr,
kronberger@hep.itp.tuwien.ac.at, mschweda@tph.tuwien.ac.at,
michael.wohlgenannt@univie.ac.at}
\vspace{0.5cm}
\end{center}
\begin{abstract}
Motivated by the recent construction of a translation-invariant
renormalizable  {\nc} model for a scalar
field~\cite{Rivasseau:2008a}, we introduce models for
{\nc} $U(1)$ gauge fields along the same lines.
More precisely, we include
some extra terms into the action with the aim of getting
rid of the UV/IR mixing.
\end{abstract}

\thispagestyle{empty}

\newpage

\section{Introduction}

Non-commuting  space-time coordinates naturally appear in various approaches
to quantum gravity, e.g. see the reviews~\cite{Douglas:2001}.
Field theories on {\nc} space generally suffer
from a new class of problematic infrared divergences which have
the same degree as the usual ultraviolet
divergences at the perturbative level.
This phenomenon is commonly referred to as {\uim},
see~\cite{Douglas:2001} and references therein.
Recently, this problem could be overcome within certain models of
scalar field theories. The first of these models, which was
introduced by Grosse and Wulkenhaar~\cite{Grosse:2003}, is the $\phi^4$ theory supplemented by an
oscillator term in the Euclidean $x$-space action: this model has been proved to be renormalizable to all orders of perturbation
theory by different methods~\cite{Grosse:2004a}. Since
the oscillator term breaks the translational invariance, Gurau, Magnen,
Rivasseau and Tanasa~\cite{Rivasseau:2008a} recently introduced
another renormalizable model in which the oscillator term in
$x$-space is replaced in the Euclidean
momentum space action by a $1/\k^2$ term (with $\k ^2 = \k^\mu \k_\mu$ and
 $\k_\mu=\th_{\mu\nu} k^\nu$, where $\th_{\mu\nu}$ are the
 non-commutativity parameters for the Euclidean space-time
 coordinates.)
This term is motivated by the fact that the 1-loop self-energy of
the standard {\nc} $\phi^4$ model has a quadratic IR divergence which is
proportional to $1/\k^2$: the new term in the  momentum space action
yields a dressed propagator at 1-loop level involving a similar
contribution.
(As a matter of fact,
such a term had already been considered earlier in connection with a
resummation procedure~\cite{Griguolo:2001,Putz:2003}.)

The deformation
matrix $( \th_{\mu\nu} )$ can be (and is) assumed to have the
simple form
\[
 ( \th_{\mu\nu} )
=\th\left(\begin{array}{cccc}
0&1&0&0\\
-1&0&0&0\\
0&0&0&1\\
0&0&-1&0
\end{array}
\right) \, ,
\qquad {\rm with} \ \; \theta \in {\bf R} \, .
\]
 The action of Gurau et al.~\cite{Rivasseau:2008a} is given in
Euclidean  momentum space by
\begin{align}\label{rivasseau-action}
\Act =\intk \left[ \inv{2} k_\mu \phi k^\mu \phi +\inv{2}m^2 \phi
\phi +\frac{a}{2}\inv{\th^2 k^2} \phi \phi +\frac{\l}{4!} \phi
\star \phi \star \phi \star \phi \right] \, ,
\end{align}
or, more explicitly~\cite{Grosse:2008a},
\[
\Act [ \hat{\phi}] =\intk \left[ \inv{2} \hat{\phi} (-k) \left(
k^2 + m^2 + \frac{a}{\th^2 k^2} \right) \hat{\phi} (k) +
\frac{\l}{4!} {\cal F} \left( \phi \star \phi \star \phi \star
\phi \right) (k) \right] \, ,
\]
where $a>0$ and where $\hat{\phi} \equiv  {\cal F} \phi$ denotes
the Fourier transform of $\phi$. This leads to the improved
propagator
\begin{align}\label{scalar-prop}
G^{\phi\phi}(k)&=\inv{k^2+ m^2+\frac{a}{\th^2 k^2}} \, ,
\end{align}
which has a ``damping'' behaviour for vanishing momentum:
\[
\lim\limits_{k\to0}G^{\phi\phi}(k)=0.
\]
As in the Grosse-Wulkenhaar model, the {\uim} is avoided due to a
mixing of long and short scales. As we already mentioned, the
model defined by the action (\ref{rivasseau-action}) is
translation-invariant and it has been proved
to be renormalizable to all orders~\cite{Rivasseau:2008a}.

Before proceeding further, we briefly spell out how the term
$1/k^2$ looks like in $x$-space. In 4 dimensions, the function
$1/x^2$ is invariant under Fourier transformation (up to a
factor), hence the term $1/k^2$ in the action
(\ref{rivasseau-action}) can be rewritten as
a non-local term:
\begin{equation}
\intk \, \hat{\phi} (-k)  \frac{1}{k^2} \hat{\phi} (k) \propto
\int d^4x \int d^4 x^{\prime} \, \phi (x) \frac{1}{(x- x^{\prime}
)^2} \phi (x^{\prime}) \equiv \int d^4 x  \, \phi
\frac{1}{\square} \phi \, .
\end{equation}
Here, the last expression is the usual short-hand notation used in
physics, where the symbol $1/\square$ denotes the Green function
$G$ associated to the differential operator $\square \equiv
\partial^{\mu} \partial_{\mu} = \partial_1^2 + \cdots +
\partial_4^2$:
\begin{equation}
\label{green}
 \square G (x) = \delta^{(4)} (x) \, , \qquad {\rm
with} \ \; G(x) = \frac{\text{const.}}{x^2} \, .
\end{equation}

As expected, matters are more complicated in gauge field theories.
Although there have been several suggestions as to how to handle the
{\uim}~\cite{Slavnov:2003,Wulkenhaar:2007}, the corresponding
models have some drawbacks. The one introduced by
Slavnov~\cite{Slavnov:2003} relies on a constraint
which reduces the degrees of freedom of the gauge field whereas
the ones involving an oscillator-type term~\cite{Wulkenhaar:2007}
(in analogy to the scalar field model of Grosse and Wulkenhaar) break the translational invariance. Accordingly, the goal of the present
letter is to put forward some ideas for generalizing
the procedure of Gurau et al.  in view of constructing a
renormalizable and translation-invariant model for $U(1)$ gauge
fields in 4 dimensional {\nc} Euclidean space.

\section{New gauge field model I}

The quadratic IR divergence of a {\nc} $U(1)$ gauge theory is
known to be of the form
\begin{align}
\Pi^{\text{IR}}_{\mu\nu}\propto\frac{\k_\mu \k_\nu}{(\k^2)^2}\,,
\end{align}
and to be independent of the chosen gauge fixing (see
references~\cite{Hayakawa:1999}).
This expression motivated the authors of reference~\cite{Putz:2003} to
introduce the following gauge invariant term into their action (in
connection with a resummation procedure):
\begin{align}
\int d^4x \, \F\star\inv{(\D^2)^2}\star\F \, .
\end{align}
Here,
\begin{align}
\F&=\th^{\mu\nu}F_{\mu\nu} \, , \qquad \ \ {\rm with} \ \;
F_{\mu\nu}=\partial_\mu A_\nu-\partial_\nu A_\mu
-\ri g\starco{A_\mu}{A_\nu}
\, ,
\nonumber\\
\D ^2 &= \D^{\mu} \star \D_\mu \, ,
\qquad {\rm with} \ \;
\D_\mu=\th_{\mu\nu}D^\nu
\,,
\end{align}
hence $\inv{\D^2}\star\F = \frac{1}{\theta^2}
\inv{D^2}\star\F$. The expression $\inv{D^2}\star\F\equiv Y$ is to
be understood as a formal power series in the gauge field $A_\mu$
which may be determined recursively as follows. First, note that
\begin{align}
\label{pre-recursion} \F &= D^2 \star \inv{D^2} \star \F =D^2 Y =
\pa^{\mu} (D_{\mu} Y)- \ig \starco{A^{\mu}}{D_{\mu}Y}
\nonumber \\
&=\square Y - \ig \pa^{\mu} \starco{A_{\mu}}{Y} - \ig
\starco{A^{\mu}}{\pa_{\mu} Y} +(\ig)^2
\starco{A^{\mu}}{\starco{A_{\mu}}{Y}}\,.
\end{align}
By applying $\inv{\square}\equiv\square^{-1}$ (i.e. the Green
function of the operator $\square$, see equation (\ref{green})) to
this relation, we find
\begin{align}\label{recursion}
 Y=\inv{\square}\F
+\frac{\ig}{\square}\pa^\mu\starco{{A}_\mu}{Y}
+\frac{\ig}{\square}\starco{{A}^\mu}{\pa_\mu Y}
-\frac{(\ig)^2}{\square}\starco{{A}^\mu}{\starco{{A}_\mu}{Y}} \,.
\end{align}
The quantity $Y$ can be determined from this equation up to an
arbitrary order:
\begin{align}
Y^{(0)}&=\inv{\square}\F\,,\nonumber\\
Y^{(1)}&=\inv{\square}\F
+\frac{\ig}{\square}\pa^\mu\starco{{A}_\mu}{\inv{\square}\F}
+\frac{\ig}{\square}\starco{{A}^\mu}{\pa_\mu \inv{\square}\F}
-\frac{(\ig)^2}{\square}\starco{{A}^\mu}{\starco{{A}_\mu}{\inv{\square}\F}}
\,,
\end{align}
and so on.

Next, we define the BRST transformations of the gauge
field $A_{\mu}$, the ghost $c$, the anti-ghost $\bar c$ and the
Lagrange multiplier $B$ as usual:
\begin{align}
\label{brst} &sA_\mu=D_\mu c\equiv \partial_\mu
c-\ig\starco{A_\mu}{c}, && s\bc=B ,
\nonumber\\
&sc=\ig{c}\star{c}, && sB=0,
\nonumber\\
& s^2\varphi=0
\quad \mbox{for} \ \; \varphi\in\left\{A_\mu,c,\bc, B \right\}
\, .
\end{align}
The $s$-variation of $A_{\mu}$ implies
$s\F=\ig\starco{c}{\F}$, from which it follows (as we will now
show) that
\begin{align}
\label{brstf} s\left( \inv{\D^2}\star\F \right)
=\ig\starco{c}{\inv{\D^2}\star\F}\,.
\end{align}
Indeed, for a field $\Phi$ transforming as $\tilde F$, i.e.
\begin{align}
\label{sphi}
s\Phi=\ig\starco{c}{\Phi}
\, ,
\end{align}
the field $D^2 \Phi$ also transforms covariantly:
 $s(D^2 \Phi) =\ig\starco{c}{D^2 \Phi}$. From
\[
s(D^2 \Phi ) = (sD^2) \, \Phi + D^2 (s \Phi )
\]
and the previous transformation law, we obtain the operatorial relation
\begin{align}
\label{sd2}
(sD^2) \bullet = -\ig  \starco{D^2 c}{\bullet}
- 2 \ig \starco{D^{\mu} c}{D_{\mu} \bullet }
\, .
\end{align}
By applying the $s$-operator to $\Phi = D^2\star\inv{D^2} \star \Phi$,
\[
s \Phi 
= (sD^2) \star \left( \inv{D^2} \star \Phi \right) +
D^2 \star s \left( \inv{D^2} \star \Phi \right)
\, ,
\]
we can deduce the transformation law of
$\inv{D^2} \star \Phi$:
\[
s \left( \inv{D^2} \star \Phi \right) = \inv{D^2} \star (s\Phi)
- \inv{D^2} \star (sD^2) \star \left( \inv{D^2} \star \Phi \right)
\, .
\]
Substitution of (\ref{sphi}) and  (\ref{sd2}) into this relation
leads to the conclusion that $\inv{D^2} \star \Phi$ transforms in
the same manner as $\Phi$,
\begin{align}
\label{sd-2} s \left( \inv{D^2} \star \Phi \right) =\ig
\starco{c}{\inv{D^2} \star \Phi } \,,
\end{align}
whence the result (\ref{brstf}).

Consider now the following action for the $U(1)$ gauge field $A_{\mu}$ in 4 dimensional {\nc} Euclidean space:
\begin{align}
\label{new-action-I}
\Gam&=\Act_{\text{inv}}+\Act_{\text{gf}}
\,,
\nonumber\\
\Act_{\text{inv}}&=\intx\left[\inv{4}F^{\mu\nu}\star F_{\mu\nu}
+\frac{\beta}{4}
\left(\inv{\D^2}\star\F\right)\star\left(\inv{\D^2}\star\F\right)\right]
\,,
\nonumber\\
\Act_{\text{gf}}&=
s \intx \, \bc \star \Bigg[
\left( 1+\frac{\gamma}{\square\wsq} \right) \partial^{\mu} A_{\mu}
- \inv{2}  B
\Bigg]
\nonumber\\
&= \intx\Bigg[B\star
\left(1+\frac{\gamma}{\square\wsq}\right)\partial^\mu A_\mu
-\inv{2}B\star B
-\bc\star\left(1+\frac{\gamma}{\square\wsq}\right)\partial^\mu D_\mu c
\Bigg]\,.
\end{align}
Here, $\beta$ and $\gamma$ are constants, and the term
parametrized by $\gamma$ has been introduced in order to improve
the IR behaviour in the ghost sector. (For $\gamma\to0$, one
recovers the Feynman gauge expression.) 
Furthermore, $\wsq= \wpart^\mu \wpart_\mu$ 
and $\wpart_\mu=\th_{\mu\nu}\partial^{\nu}$.

The action $\Gam$ is invariant under the BRST transformations
(\ref{brst}), (\ref{brstf}). Its bilinear part $\Act_{\text{bil}}$
yields the following equations of motion for the free fields:
\begin{align}
0 &= \var{\Act_{\text{bil}}}{A^\nu}
=-\left(\square\d_{\nu\mu} -\partial_\nu\partial_\mu\right)A^\mu
+\frac{\beta}{\wsq^2}\wpart_\nu\wpart_\mu A^\mu
-\left(1+\frac{\gamma}{\square\wsq}\right)\partial_\nu B \,,
\nonumber\\
0 &=\var{\Act_{\text{bil}}}{B}
=\left(1+\frac{\gamma}{\square\wsq}\right)\partial^\mu A_\mu-B \,,
\nonumber\\
0 &=\var{\Act_{\text{bil}}}{\bc}
=-\left(1+\frac{\gamma}{\square\wsq}\right)\square c \,.
\end{align}
This leads to the following propagators in momentum space:
\begin{align}
G^{A}_{\mu\nu}(k)&=\inv{k^2}\left(\d_{\mu\nu}
+\frac{k_\mu k_\nu}{k^2}
-\frac{k_\mu k_\nu}{k^2\left(1+\frac{\gamma}{k^2\k^2}\right)^2}
-\beta\frac{\k_\mu\k_\nu}{(\k^2)^2(k^2+\frac{\beta}{\k^2})}\right)
\,,
\nonumber\\
G^{\bc c}(k)&=\inv{k^2+\frac{\gamma}{\k^2}}
\, .
\end{align}
Since the gauge field propagator  $G^{A}_{\mu\nu}$ involves an
overall factor $\inv{k^2}$, it is not damped for $k\to 0$ and one
may argue that it does not sufficiently mix long and short scales.
If one takes this issue of ``mixing'' more seriously, one is led
to the alternative model presented in the next section.

\section{New gauge field model II}
Considering that the scaling behaviour of the propagator
(\ref{scalar-prop}) of Gurau et al.~\cite{Rivasseau:2008a}
ensures the IR finiteness of their model, we look for a BRST
invariant action leading to a similar propagator for the $U(1)$
gauge field $A_{\mu}$. Accordingly, we introduce the following
action in 4 dimensional {\nc} Euclidean space:
\begin{align}\label{new-action-II}
\Gam&=\Act_{\text{inv}}
+\Act_{\text{gf}}\,,
\nonumber\\
\Act_{\text{inv}}&=\intx\left[\inv{4}F^{\mu\nu}\star F_{\mu\nu}
+\inv{4}F^{\mu\nu}\star\inv{D^2\D^2}\star F_{\mu\nu}\right]
\,,
\nonumber\\
\Act_{\text{gf}}&=
s \intx \, \bc \star \Bigg[
\left( 1+\frac{1}{\square\wsq} \right) \partial^{\mu} A_{\mu}
- \frac{\alpha}{2}  B
\Bigg]
\nonumber\\
&=\intx\Bigg[B\star
\left(1+\inv{\square\wsq}\right)\partial^\mu A_\mu-\frac{\alpha}{2}B\star B
-\bc\star\left(1+\inv{\square\wsq}\right)\partial^\mu D_{\mu} c
\Bigg]
\, .
\end{align}
Here, $\alpha$ is a real parameter and $\inv{D^2\D^2}\star F_{\mu\nu}$ is again to be understood as a formal power series in the gauge field $A_\mu$. The functional $\Gam$ is invariant under the BRST transformations (\ref{brst}) which imply
\begin{align}
s\left(\inv{D^2\D^2}\star F_{\mu\nu}\right)
=\ig\starco{c}{\inv{D^2\D^2}\star F_{\mu\nu}}
\, .
\end{align}
The bilinear part of the action now leads to the following
equations of motion for the free fields:
\begin{align}
0 &=\var{\Act_{\text{bil}}}{A^\nu}
=-\left(1+\inv{\square\wsq}\right)\left(\square\d_{\nu\mu}
-\partial_\nu\partial_\mu\right)A^\mu
-\left(1+\inv{\square\wsq}\right)\partial_\nu B \,,
\nonumber\\
 0 &=  \var{\Act_{\text{bil}}}{B}
=\left(1+\inv{\square\wsq}\right)\partial^\mu A_\mu -\alpha B \,,
\nonumber\\
 0 &=  \var{\Act_{\text{bil}}}{\bc}
=-\left(1+\inv{\square\wsq}\right)\square c \,.
\end{align}
Hence, we get the following propagators in momentum space:
\begin{align}
G^{A}_{\mu\nu}(k)&=\inv{k^2+\inv{\k^2}}\left(-\d_{\mu\nu}
+\frac{k_\mu k_\nu}{k^2}-\alpha\frac{k_\mu k_\nu}{k^2+\inv{\k^2}}\right)\,,
\nonumber\\
G^{\bc c}(k)&=\inv{k^2+\inv{\k^2}}
\,.
\end{align}
If one chooses the Landau gauge $\alpha=0$ for the gauge
parameter, then the gauge field  propagator simplifies to
\begin{align}
G^{A}_{\mu\nu}(k)&=\inv{k^2+\inv{\k^2}}\left(-\d_{\mu\nu}
+\frac{k_\mu k_\nu}{k^2}\right)
\,.
\end{align}

\section{Concluding remarks}

In the preceding sections, we introduced two natural
models for non-commutative $U(1)$ gauge fields.
These models are both BRST-invariant and translation-in\-vari\-ant, 
and they are devised for curing the {\uim} problem. 
The second model has the advantage that the gauge field
propagator has an improved ``damping'' behaviour for vanishing
momentum. The question whether this property is sufficient for 
ensuring the renormalizability of the model obviously requires
further and more involved investigations (work in progress).

\subsection*{Acknowledgments}
It is a great pleasure to thank Jean-Christophe Wallet for his
valuable comments on a preliminary version of the present paper.

The work of D.~N.~Blaschke, E.~Kronberger and M.~Wohlgenannt 
was supported by the ``Fonds zur F\"orderung 
der Wissenschaftlichen Forschung'' (FWF) 
under contracts P20507-N16, P19513-N16 and P20017-N16, respectively.


\end{document}